\documentclass[a4paper,12pt]{article}

\usepackage{graphics}

\newcommand{\ket}[1]{\ensuremath{\left| #1 \right\rangle}}

\newcommand{\br}[1]{\ensuremath{\left\langle #1 \right.}}

\newcommand{\bra}[1]{\ensuremath{\left. \br{#1} \right|}}

\newcommand{\bk}[2]{\br{{#1}}\ket{{#2}}}

\newcommand{\kb}[2]{\ket{{#1}}\bra{{#2}}}

\newcommand{\ip}[3]{\bra{{#1}} {#2} \ket{{#3}}}

\newcommand{\proj}[1]{\kb{{#1}}{{#1}}}

\newcommand{\trace}[1]{\ensuremath{\mathrm{Tr}\left[{#1}\right]}}

\newcommand{\partrace}[2]{\ensuremath{\mathrm{Tr}_{{#2}}\left[{#1}\right]}}

\newcommand{\mean}[1]{\ensuremath{\left\langle
                                    {#1} \right\rangle}}

\newcommand{\magn}[1]{\ensuremath{\left| {#1} \right|^2}}

\newcommand{\com}[2]{\ensuremath{\left[{#1},{#2}\right]_-}}
\newcommand{\acom}[2]{\ensuremath{\left[{#1},{#2}\right]_+}}

\newcommand{\partime}[1]{\ensuremath{
    \frac{\partial {#1}}{\partial t}}}

\newcommand{\pictpict}[4]
{
\begin{figure}[htb]
    \resizebox{\textwidth}{!}{
        \includegraphics{#1}
        \includegraphics{#2}
        }
    \caption{#3\label{fg:#4}}
\end{figure}
}

\begin{document}

\title{The density matrix in the de Broglie-Bohm approach}
\author{O J E Maroney\\
University of Bristol\\
HH Wills Physics Laboratory\\
Bristol\\
BS8 1TL\\
o.maroney@bristol.ac.uk}
\date{ }

\maketitle

If the density matrix is treated as an objective description of
individual systems, it may become possible to attribute the same
objective significance to statistical mechanical properties, such
as entropy or temperature, as to properties such as mass or
energy. It is shown that the de Broglie-Bohm interpretation of
quantum theory can be consistently applied to density matrices as
a description of individual systems. The resultant trajectories
are examined for the case of the delayed choice interferometer,
for which Bell\cite{Bel80} appears to suggest that such an
interpretation is not possible. Bell's argument is shown to be
based upon a different understanding of the density matrix to that
proposed here.

\section{INTRODUCTION}
\label{introduction} In quantum theory, it is normally assumed
that an individual system is described by a pure state
$\ket{\Phi}$. The density matrix can then enter into the theory in
one of two ways:
\begin{enumerate}
\item The system is composed of entangled subsystems.
Tracing over the Hilbert space of a subsystem gives a density
matrix which contains the statistics of measurements performed
only on the remainder of the system:
\begin{equation}
\rho(Y)=\partrace{\proj{\Phi(X,Y)}}{X}. \label{eq:reduced}
\end{equation}

\item The system is not prepared as an ensemble of identical states, but occurs in an ensemble for which
the state $\ket{\Phi_i}$ appears randomly with probability $p_i$:
\begin{equation}
\rho=\sum_i p_i \proj{\Phi_i}.
\end{equation}
\end{enumerate}

In both cases, the properties of the density matrix have
significant differences from an equivalent classical probability
distribution.  For entangled states, the entropy of a subsystem
can exceed the entropy of the overall state, which cannot happen
classically, while for
 a statistical ensemble, the density matrix does not
uniquely identify an underlying distribution of pure states.

\begin{table}[h]
    \begin{tabular}{cccc}
        \multicolumn{4}{l}{Ensemble 1} \\
        \multicolumn{2}{c}{$\rho_0 = \proj{0}$} & \multicolumn{2}{c}{$\rho_1 = \proj{1}$} \\
        \multicolumn{2}{c}{$p_0 = \frac12$} & \multicolumn{2}{c}{ $p_1=\frac12$} \\
        \multicolumn{4}{l} {Ensemble 2} \\
        \multicolumn{2}{c}{$\rho_A = \proj{u}$} & \multicolumn{2}{c}{$\rho_B = \proj{v}$} \\
        \multicolumn{2}{c}{$p_A = \frac12$} & \multicolumn{2}{c}{$p_B = \frac12$} \\
        \multicolumn{2}{c}{$\ket{u}=\frac1{\sqrt{2}}\left(\ket{0}+\ket{1}\right)$}
              & \multicolumn{2}{c}{$\ket{v}=\frac1{\sqrt{2}}\left(\ket{0}-\ket{1}\right)$} \\
        \multicolumn{4}{l}{Ensemble 3} \\
        $\rho_0 = \proj{0}$ & $\rho_1 = \proj{1}$ & $\rho_A = \proj{u} $ & $\rho_B = \proj{v}$ \\
        $p_0 = \frac14 $ & $p_1 = \frac14 $ & $p_A = \frac14 $ &
        $p_B = \frac14$
    \end{tabular}
\begin{equation}\label{rho:ensemble1}\end{equation}
\end{table}

For example, the ensembles defined upon a spin-$\frac{1}{2}$
system, in (\ref{rho:ensemble1}), all produce the density matrix
$\rho=\frac12 I$, where $I$ is the identity.  All the statistical
outcomes of an experiment can be calculated directly from the
density matrix, so there is no possible measurement procedure that
can distinguish between these three ensembles.

It has been suggested that the density matrix should be treated as
a description of a single quantum system rather than as ensemble
of systems (see, for example \cite{AA98}).  Let us be explicit as
to what is meant by this.  In Ensemble 1 above, the system is in
the state $\ket{0}$ half the time.  If an observable $\proj{A}$ is
measured on a system, the probability of observing the outcome $a$
is
\begin{equation}
p(a|0)=\magn{\bk{0}{A}}.
\end{equation}
Similarly, for systems in the state $\ket{1}$ the probability is
$p(a|1)$.  It is only over the ensemble of states that the result
\begin{equation}
p(a)=p_0 p(a|0)+p_1 p(a|1)
\end{equation}
is obtained.

If the individual state were described by the density matrix
$\rho$, then for every system the probability of outcome $a$
occurring would be
\begin{equation}
p(a)=\trace{\rho \proj{A}}.
\end{equation}
This will not, in general, be equal to either $p(a|0)$ or
$p(a|1)$.  So the question is
 whether the potential response to a measurement of an individual system {\em must} be described
by a pure state.  To distinguish these cases, we will use the
symbol $\varrho$ to represent the individual state density matrix,
and $\rho$ to represent a statistical ensemble from now on.

As long as the states $\proj{0}$ and $\proj{1}$ occur randomly
with probabilities $p_0$ and $p_1$, it is impossible to establish,
by experimental means, any difference between the statistical
ensemble constructed out of such pure states, and an ensemble of
states where the statistical outcomes of measurements upon every
individual system is given by probabilities $p(a)$ that come from
a density matrix.  This fact alone provides grounds for arguing
that it is unreasonable to require, as a matter of principle, that
individual systems be described by pure states, rather than
density matrices.  The non-uniqueness of the decomposition of the
density matrix adds the further complication that, even if we are
to assume an ensemble of systems is constructed from individual
pure states, we are unable to determine, by experimental means,
which set of pure states is involved\footnote{There are, of
course, situations, such as in communication problems\cite{CN01},
where we are in possession of {\it a priori} knowledge of the
signal states from which a density matrix is composed.  In such
situations the statistical ensemble remains the correct view to
take.}.

A further motivation is provided by thermodynamics. The derivation
of thermodynamic quantities from the statistical mechanics of an
ensemble of large numbers of systems leads to significant
conceptual problems regarding the status of the second law of
thermodynamics (\cite{Pop57,LR90,HMZ94,Alb00,Uff01} amongst many
others).
  If the density matrix can describe the state of
an individual quantum system, then that system can have a non-zero
entropy
\begin{equation}
S=\trace{\varrho \ln [\varrho]}
\end{equation}
independantly of its belonging to a particular ensemble.  This
would be expected to have a significant affect upon the discussion
of the foundations of thermodynamics and in particular the
discussion of Maxwell's demon and fluctuation
phenomena\footnote{see \cite[Chapter 10]{Mar02} for a brief
discussion of this}.

It should be noted that this suggestion would be more difficult to
maintain for classical systems.
  The ontology of classical mechanics
describes systems possessing definite values for all properties.
The classical probability distribution uniquely defines the
underlying states and their distribution, and by a suitably
idealised measurement the particular state can be discovered,
non-destructively, in each individual case.

In the quantum case, even for pure states the possession of a
particular property can only be described through probability
distributions. Treating the density matrix as having the same
ontological status as the pure state then seems to present no
additional problems.

However, in \cite{Bel80}, Bell appears to suggest that this is not
possible for the de Broglie-Bohm interpretation:
\begin{quote}
So in the de Broglie-Bohm theory a fundamental significance is
given to the wavefunction, and it cannot be transferred to the
density matrix.  This is here illustrated for the one-particle
density matrix, but it [is] equally so for the world density
matrix if a probability distribution over world wavefunctions is
considered.  Of course the density matrix retains all its usual
practical utility in connection with quantum statistics.
\end{quote}
suggesting that only the statistical mechanics of the kind given
in \cite{BH96a} is valid for this interpretation.

In this paper we show that it is, in fact, possible to apply the
de Broglie-Bohm approach directly to the density matrix and to
construct consistent trajectory solutions for this.  We then apply
this to the delayed choice interferometer (Section
\ref{interferometer}) considered by Bell and show no special
problems arise.

\section{BOHM TRAJECTORIES FOR THE DENSITY MATRIX} \label{bohm}

The de Broglie-Bohm interpretation\cite{Boh52a,Boh52b,BH93,Hol93}
has traditionally been applied only to pure states.  Density
matrices have been treated only in the context of statistical
ensembles\cite{BH96a}. To treat the density matrix as a
description of an individual system, we will apply the formalism
developed by Brown and Hiley\cite{BH00}, who use the Bohm approach
within a purely algebraic framework.

\subsection{The Algebraic Approach}
In \cite{BH00}, it is suggested that the Bohm approach can be
generalised to the coupled algebraic equations \footnote{
\begin{eqnarray*}
\com{A}{B}&=& AB-BA \\
\acom{A}{B}&=& AB+BA
\end{eqnarray*}
}

\begin{eqnarray}
\partime{\varrho}&=&
    \imath \com{\varrho}{H}  \label{eq:qliou}\\
\varrho \partime{\hat{S}} &=&
    -\frac12 \acom{\varrho}{H} \label{eq:aqhj}.
\end{eqnarray}

Equation \ref{eq:qliou} is simply the quantum Liouville equation,
which represents the conservation of probability, and reduces to
the familiar form of
\begin{equation}
\frac{\partial R(x)^2}{\partial t}+\mathbf{\nabla \cdot j}=0
\end{equation}
where $\mathbf{j}$ is the probability current
\begin{equation}
\mathbf{j}=R(x)^2 \frac{\mathbf {\nabla} S(x)}{m}
\end{equation}
in the case where the system is in a pure state
$\varrho=\proj{\psi}$ and $\bk{x}{\psi}=R(x)e^{\imath S(x)}$.

The second equation is the algebraic generalisation of the Quantum
Hamilton-Jacobi, which reduces to
\begin{equation}
-\frac{\partial S}{\partial t}= \frac{(\nabla S)^2}{2m}+V-
\frac{\nabla ^2R}{2mR} \label{eq:qhj}
\end{equation}
for pure states.

The operator $\hat{S}$ is a phase operator, and this equation can
be taken to represent the energy of the quantum system. The
application of this to the Aharanov-Bohm, Aharanov-Casher and
Berry phase effects is demonstrated in \cite{BH00}.

\cite{BH00} are concerned with the problem of symplectic symmetry,
so their paper deals mainly with constructing momentum
representations of the Bohm trajectories, for pure states, and
does not address the issue of when the density matrix is a mixed
state. Here we will be concentrating entirely upon the mixed state
properties of the density matrix, and so we will leave aside the
questions of symplectic symmetry and the interpretation of
Equation \ref{eq:aqhj}. Instead we will assume the Bohm
trajectories are defined using a position 'hidden variable' or
'beable', and will concentrate on Equation \ref{eq:qliou}.

The Brown-Hiley method, for our purposes, can be summarised by the
use of algebraic probability currents
\begin{eqnarray}
J_X &=& \nabla_P \left(\varrho H \right) \\
J_P &=& \nabla_X \left(\varrho H \right)
\end{eqnarray}
for which
\begin{equation}
\imath \frac{\partial \varrho}{\partial t} +
    \com{J_X}{P}-\com{J_P}{X}=0.
\end{equation}
To calculate trajectories in the position representation (which
Brown and Hiley refer to as constructing a 'shadow phase space')
from this we must project out the specific location $x$, in the
same manner as we project out the wavefunction from the Dirac ket
$\psi(x)=\bk{x}{\psi}$.
\begin{equation}
\imath \partime{\ip{x}{\varrho}{x}} +
    \ip{x}{\com{J_X}{P}}{x} - \ip{x}{\com{J_P}{X}}{x}=0.
\end{equation}
The second commutator vanishes and the first commutator is
equivalent to the divergence of a probability current
\begin{equation}
\mathbf{\nabla_x \cdot J(x)} =\ip{x}{\com{J_X}{P}}{x}
\end{equation}
leading to the conservation of probability equation
\begin{equation}
\partime{P(x)}+\mathbf{\nabla_x \cdot J(x)} =0.
\end{equation}

To see the general solution to this, we will note that the density
matrix of a system will always have a diagonal basis
\ket{\phi_a}(even if this basis is not the energy eigenstates),
for which
\begin{equation}
\varrho=\sum_a w_a \proj{\phi_a}.
\end{equation}
Note, the $w_a$ are {\em not} interpreted here as statistical
weights in an ensemble. There are physical properties of the state
$\varrho$, with a similar status to the probability amplitudes in
a superposition of states.

We can put each of the basis states into the polar form
\begin{equation}
R_a(x)e^{\imath S_a(x)}=\bk{x}{\phi_a}
\end{equation}
so the probability density is just
\begin{equation}
P(x)=\sum_a w_a R_a(x)^2.
\end{equation}

The probability current now takes the more complex form
\begin{equation}
\mathbf{J}(x)=\sum_a w_a R_a(x)^2 \mathbf{\nabla} S_a(x).
\end{equation}

So far we have not left standard quantum theory\footnote{The
probability current is a standard part of quantum theory, as its
very existence is necessary to ensure the conservation of
probability.}. We may do this by now constructing trajectory
solutions $X(t)$, in the manner of the Bohm approach, by
integrating along the flow lines of this probability
current\cite{BH93,Hol93,BH00}. This leads to
\begin{equation}
m\partime{\mathbf{X}(t)}=\frac{\mathbf{J}(X(t))}{P(X(t))}=
    \frac{\sum_a w_a R_a(X(t))^2 \mathbf{\nabla} S_a(X(t))}
    {\sum_a w_a R_a(X(t))^2} \label{eq:trajrho}.
\end{equation}
Notice the important fact that, when the density matrix represents
a pure state, this reduces to {\em exactly} the Bohm
interpretation for pure states.

The most notable feature of Equation \ref{eq:trajrho} is that the
constructed particle velocity is {\em not} the statistical average
of the velocities \mean{V(t)}, that would have been calculated
from the interpretation of $\rho=\sum_a w_a \proj{\phi_a}$ as an
ensemble:
\begin{equation}
\mean{\mathbf{V}(t)}=\sum_a w_a \mathbf{\nabla} S_a(X(t)).
\end{equation}

This should not be too surprising however. We are interpreting the
entire density matrix as providing a pilot wave to guide the
individual particle motion. {\em All} the elements of the density
matrix are physically present, for a particle at $X(t)$, and each
state \ket{\phi_a} contributes a 'degree of activity', given by
$R_a(x)^2$, to the motion of the trajectory, in addition to the
weighting $w_a$. If a particular state has a probability amplitude
that is very low, in a given location, then even if its weight
$w_a$ is large, it may make very little contribution when the
trajectory passes through that location, and vice versa.

Let us consider this with the simple example of a system which has
two states \ket{\phi_a} and \ket{\phi_b}. The probability
equations are
\begin{eqnarray}
P(x) &=& w_a R_a(x)^2+w_b R_b(x)^2 \\
\mathbf{J}(x) &=& w_a R_a(x)^2 \mathbf{\nabla} S_a(x)
    +w_b R_b(x)^2 \mathbf{\nabla} S_a(x).
\end{eqnarray}
Let us suppose that the two states \ket{\phi_a} and \ket{\phi_b}
are superorthogonal\footnote{Superorthogonality is defined in
\cite{BH93} as a situation where two wavepackets are completely
non-overlapping in the configuration space of the beable:
$\Psi_a(X)\Psi_b(X) \approx 0$.  It is a much stronger condition
than orthogonality: $\bk{\Psi_a(X)}{\Psi_b(X)} = 0$.
Superorthogonality of measuring device states plays a key role in
the Bohmian resolution of the measurement problem.}. This implies
$\phi_a(X)\phi_b(X) \approx 0$ for all $X$. This must also hold
for the probability amplitudes $R_a(X)R_b(X) \approx 0$. If the
particle trajectory $X(t)$ is located in an area where $R_a(X)$ is
non-zero, then now the value of $R_b(X) \approx 0$. The
probability equations become
\begin{eqnarray}
P(X) & \approx & w_a R_a(X)^2 \\
\mathbf{J}(X) & \approx & w_a R_a(X)^2 \mathbf{\nabla} S_a(X)
\end{eqnarray}
and so the particle trajectory
\begin{equation}
m \partime{\mathbf{X}(t)} \approx \mathbf{\nabla} S_a(X(t))
\end{equation}
follows the path it would have taken if system was in the pure
state \ket{\phi_a}. In this situation, where there is no overlap
between the states, then the Bohm trajectories behave in exactly
the same manner as if the system had, in fact, been in a
statistical ensemble.

Now, if we make the assumption necessary to the Bohm
interpretation, that the initial co-ordinate of the particle
trajectory occurs at position $X(0)$, with a probability given by
$P(X(0))$, it is apparent that the trajectories, at time $t$ will
be distributed at positions $X(t)$ with probability $P(X(t))$. We
have therefore consistently extended the Bohm approach to treat
density matrices (and therefore thermal states) as a fundamental
property of individual systems, rather than statistical ensembles.
As we know that the statistics of the outcomes of experiments can
be expressed entirely in terms of the density matrix, we also know
that the results of any measurements in the approach will exactly
reproduce all the statistical results of standard quantum theory.

\section{CORRELATIONS AND MEASUREMENT}

We will now look at how this extension of the Bohm interpretation
affects the discussion of correlations and measurements.

The general state of a quantum system consisting of two subsystems
will be a joint density matrix $\varrho_{1,2}$. This {\em joint}
density matrix must be diagonalised, before we project onto the
configuration space of {\em both} particle positions, using
\ket{x_1,x_2}. We can represent this projection by a 6 dimensional
vector, $x$, in the configuration space, incorporating the 3
dimensions of $x_1$ and the 3 dimensions of $x_2$. The probability
equations are simply
\begin{eqnarray}
P(x_1,x_2) &=& \sum_a w_a R_a(x_1,x_2)^2 \\
\mathbf{J}(x_1,x_2) &=&
    \sum_a w_a R_a(x_1,x_2)^2 \mathbf{\nabla_x} S_a(x_1,x_2).
\end{eqnarray}
The probability current can be divided into two
\begin{equation}
\mathbf{J}(x_1,x_2) = \mathbf{J_1}(x_1,x_2)+\mathbf{J_2}(x_1,x_2)
\end{equation}
where
\begin{eqnarray}
\mathbf{J_1}(x_1,x_2) &=& \sum_a w_a R_a(x_1,x_2)^2
    \mathbf{\nabla_{x_1}} S_a(x_1,x_2) \\
\mathbf{J_2}(x_1,x_2) &=& \sum_a w_a R_a(x_1,x_2)^2
    \mathbf{\nabla_{x_2}} S_a(x_1,x_2).
\end{eqnarray}
The conservation of probability is expressed as
\begin{equation}
\partime{P(x_1,x_2)}+\mathbf{\nabla_{x_1} \cdot J}(x_1,x_2)
    +\mathbf{\nabla_{x_2} \cdot J}(x_1,x_2)=0.
\end{equation}
The particle trajectories must be described by a joint co-ordinate
$X(t)$ in the configuration space of both particles, which evolves
according to
\begin{equation}
m\partime{\mathbf{X}(t)}=\frac{\mathbf{J}(X(t))}{P(X(t))}.
\end{equation}
If we separate this into the trajectories of the two separate
particles $X_1(t)$ and $X_2(t)$, this becomes the coupled
equations
\begin{eqnarray}
m\partime{\mathbf{X_1}(t)} &=&
    \frac{\mathbf{J_1}(X_1(t),X_2(t))}{P(X_1(t),X_2(t))} \\
m\partime{\mathbf{X_2}(t)} &=&
    \frac{\mathbf{J_2}(X_1(t),X_2(t))}{P(X_1(t),X_2(t))}.
\end{eqnarray}
We see, exactly as in the pure state situation, that the evolution
of one particle trajectory is dependant upon the instantaneous
location of the second particle, and vice versa.

The first special case to consider is when the density matrices
are uncorrelated
\begin{equation}
\varrho_{1,2}=\varrho_1 \otimes \varrho_2.
\end{equation}
The probability equations reduce to the form
\begin{eqnarray}
P(x_1,x_2)=P(x_1)P(x_2)=
    \sum_a w_a R_a(x_1)^2 \sum_b w_b R_b(x_2)^2 \\
\mathbf{J}(x_1,x_2)=P(x_2)\mathbf{J_1}(x_1)
    +P(x_1)\mathbf{J_2}(x_2)
\end{eqnarray}
where
\begin{eqnarray}
\mathbf{J_1}(x_1) &=&
    \sum_a w_a R_a(x_1)^2 \mathbf{\nabla_{x_1}} S_a(x_1) \\
\mathbf{J_2}(x_2) &=&
    \sum_b w_b R_b(x_2)^2 \mathbf{\nabla_{x_2}} S_b(x_2).
\end{eqnarray}
The resulting trajectories
\begin{eqnarray}
m\partime{\mathbf{X_1}(t)} &=&
    \frac{\mathbf{J_1}(X_1(t))}{P(X_1(t))} \\
m\partime{\mathbf{X_2}(t)} &=&
    \frac{\mathbf{J_2}(X_2(t))}{P(X_2(t))}
\end{eqnarray}
show the behaviour of the two systems are completely independant.

Now let us consider a correlated density matrix
\begin{equation}
\varrho_{1,2}=\frac12 \left(
    \proj{\phi_a\chi_a} +\proj{\phi_b\chi_b} \right)
\end{equation}
where the \ket{\phi} states are for system 1 and the \ket{\chi}
states are for system 2. The polar decompositions
\begin{eqnarray}
R_a(x_1)R_a(x_2)e^{\imath S_a(x_1)+S_a(x_2)}&=&
    \bk{x_1,x_2}{\phi_a\chi_a} \\
R_b(x_1)R_b(x_2)e^{\imath S_b(x_1)+S_b(x_2)}&=&
    \bk{x_1,x_2}{\phi_b\chi_b}
\end{eqnarray}
lead to probability equations
\begin{eqnarray}
P(x_1,x_2) &=& \frac12 \left(R_a(x_1)^2 R_a(x_2)^2
    +R_b(x_1)^2 R_b(x_2)^2\right) \\
\mathbf{J}(x_1,x_2) &=& \frac12 \left(R_a(x_1)^2 R_a(x_2)^2
    (\mathbf{\nabla_{x_1}} S_a(x_1)+\mathbf{\nabla_{x_2}} S_b(x_2))\right. \nonumber\\
&&  +\left. R_b(x_1)^2 R_b(x_2)^2
        (\mathbf{\nabla_{x_1}} S_b(x_1)+\mathbf{\nabla_{x_2}}
        S_b(x_2))\right).
\end{eqnarray}
The trajectories, $X(t)$, are then given by
\begin{equation}
m\partime{\mathbf{X_1}(t)} =
        \frac{R_a(X_1(t))^2 R_a(X_2(t))^2 \mathbf{\nabla_{X_1}} S_a(X_1(t))
            +R_b(X_1(t))^2 R_b(X_2(t))^2 \mathbf{\nabla_{X_1}} S_b(X_1(t))}
            {R_a(X_1(t))^2 R_a(X_2(t))^2+R_b(X_1(t))^2 R_b(X_2(t))^2}
\end{equation}

\begin{equation}
    m\partime{\mathbf{X_2}(t)} =
        \frac{R_a(X_1(t))^2 R_a(X_2(t))^2 \mathbf{\nabla_{X_2}} S_a(X_2(t))
            +R_b(X_1(t))^2 R_b(X_2(t))^2 \mathbf{\nabla_{X_2}} S_b(X_2(t))}
            {R_a(X_1(t))^2 R_a(X_2(t))^2+R_b(X_1(t))^2
            R_b(X_2(t))^2}.
\end{equation}
Now in general this will lead to a complex coupled behaviour.
However, if {\em either} of the states \ket{\phi} or \ket{\chi}
are superorthogonal, then relevant co-ordinate, $X_1$ or $X_2$
respectively, will be active for only one of the $R_a$ or $R_b$
states. For example, suppose the \ket{\chi} states are
superorthogonal
\begin{equation}
R_a(X_2)R_b(X_2) \approx 0.
\end{equation}
For a given location of $X_2$, only one of these probability
densities will be non-zero. If we suppose this is the \ket{\chi_a}
wavepacket, then $R_b(X_2)^2 \approx 0$. The trajectory equations
become
\begin{eqnarray}
m\partime{\mathbf{X_1}(t)} &=&
    \frac{R_a(X_1(t))^2 R_a(X_2(t))^2 \mathbf{\nabla_{X_1}} S_a(X_1(t))}
         {R_a(X_1(t))^2 R_a(X_2(t))^2} \nonumber \\
         &=& \mathbf{\nabla_{X_1}} S_a(X_1(t)) \\
m\partime{\mathbf{X_2}(t)} &=&
    \frac{R_a(X_1(t))^2 R_a(X_2(t))^2 \mathbf{\nabla_{X_2}} S_a(X_2(t))}
        {R_a(X_1(t))^2 R_a(X_2(t))^2} \nonumber \\
        &=&\mathbf{\nabla_{X_2}} S_a(X_2(t)).
\end{eqnarray}
Both trajectories behave as if the system was in the pure state
\ket{\phi_a\chi_a}. If the location of $X_2$ had been within the
\ket{\chi_b} wavepacket, then the trajectories would behave
exactly as if the system were in the pure state
\ket{\phi_b\chi_b}. The trajectories, as a whole, behave as if the
system was in a statistical mixture of states, as long as at least
one of the subsystems has superorthogonal states.

The loss of phase coherence does not play a fundamental role in
the Bohm theory of measurement. It is the superorthogonality that
is important. This carries directly over into the density matrix
description. It is a simple matter to generalise the above
arguments to a general N-body system, or to consider states where
the diagonalised density matrix involves entangled states.

\section{THE DELAYED CHOICE INTERFEROMETER}\label{interferometer}

We will now consider the Bohm trajectories for the delayed choice
version of the two slit experiment.  The basic experimental
arrangement is shown in Figure \ref{fg:interf}.

\pictpict{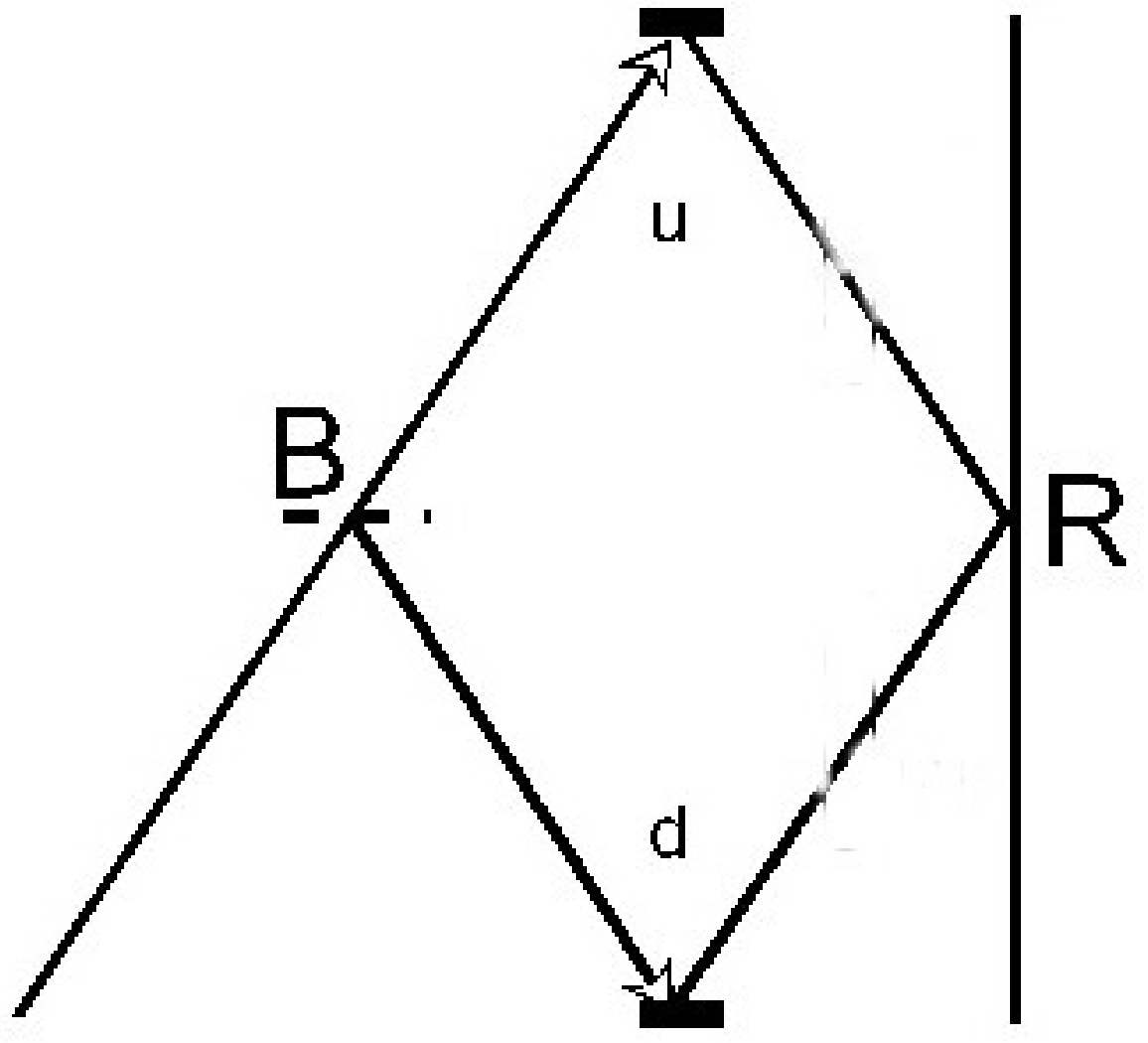}{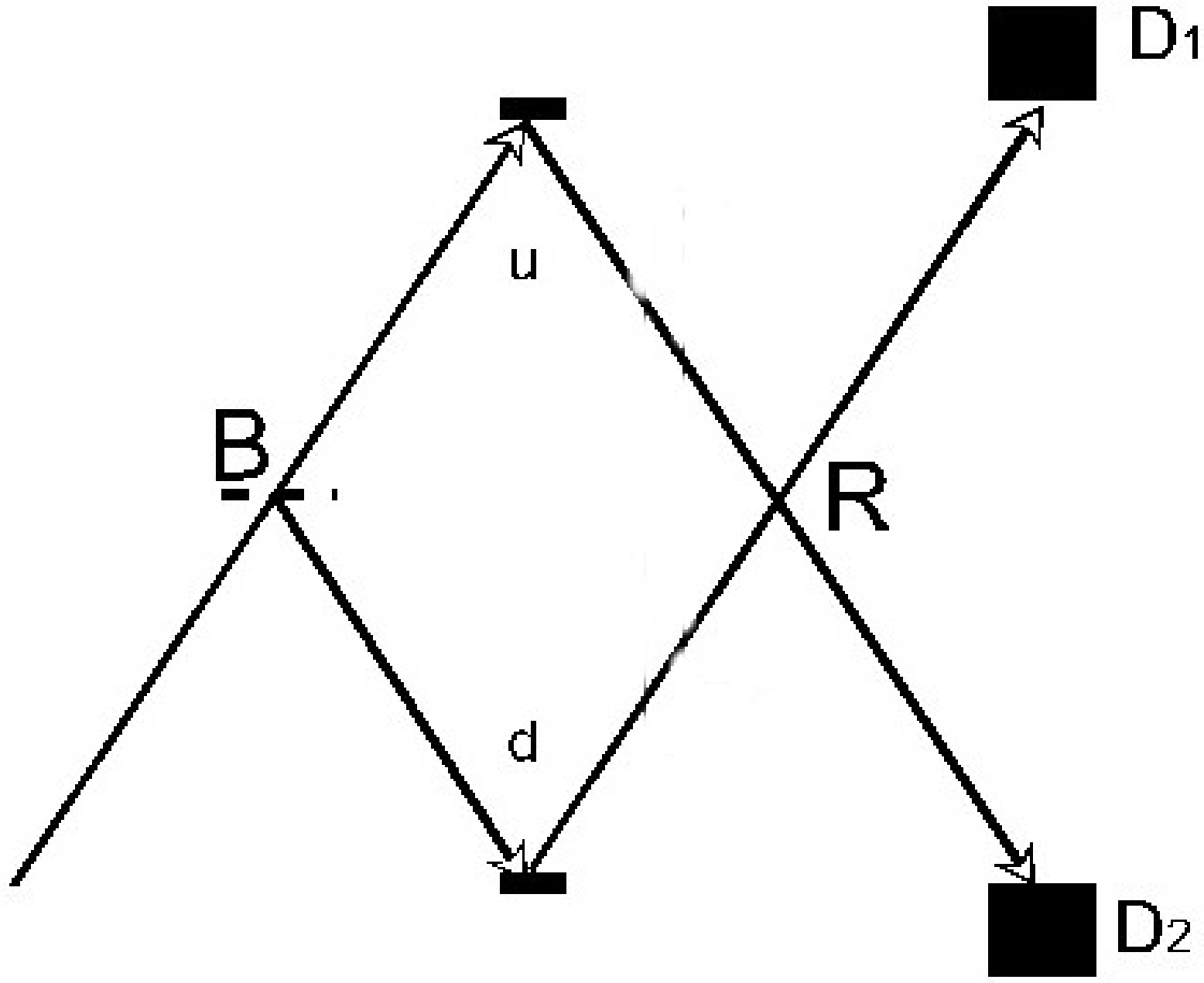} {Delayed Choice Interferometer:
interference (left); "which path" (right)}{interf}

The Bohm trajectories for pure states in this interferometer have
been discussed extensively
\cite{ESSW92,DHS93,DFGZ93,ESSW93,AV96,Cun98,Scu98,CHM00,Mar02}, in
the context of the claim in \cite{ESSW92} that the trajectories
are "surrealistic".

\pictpict{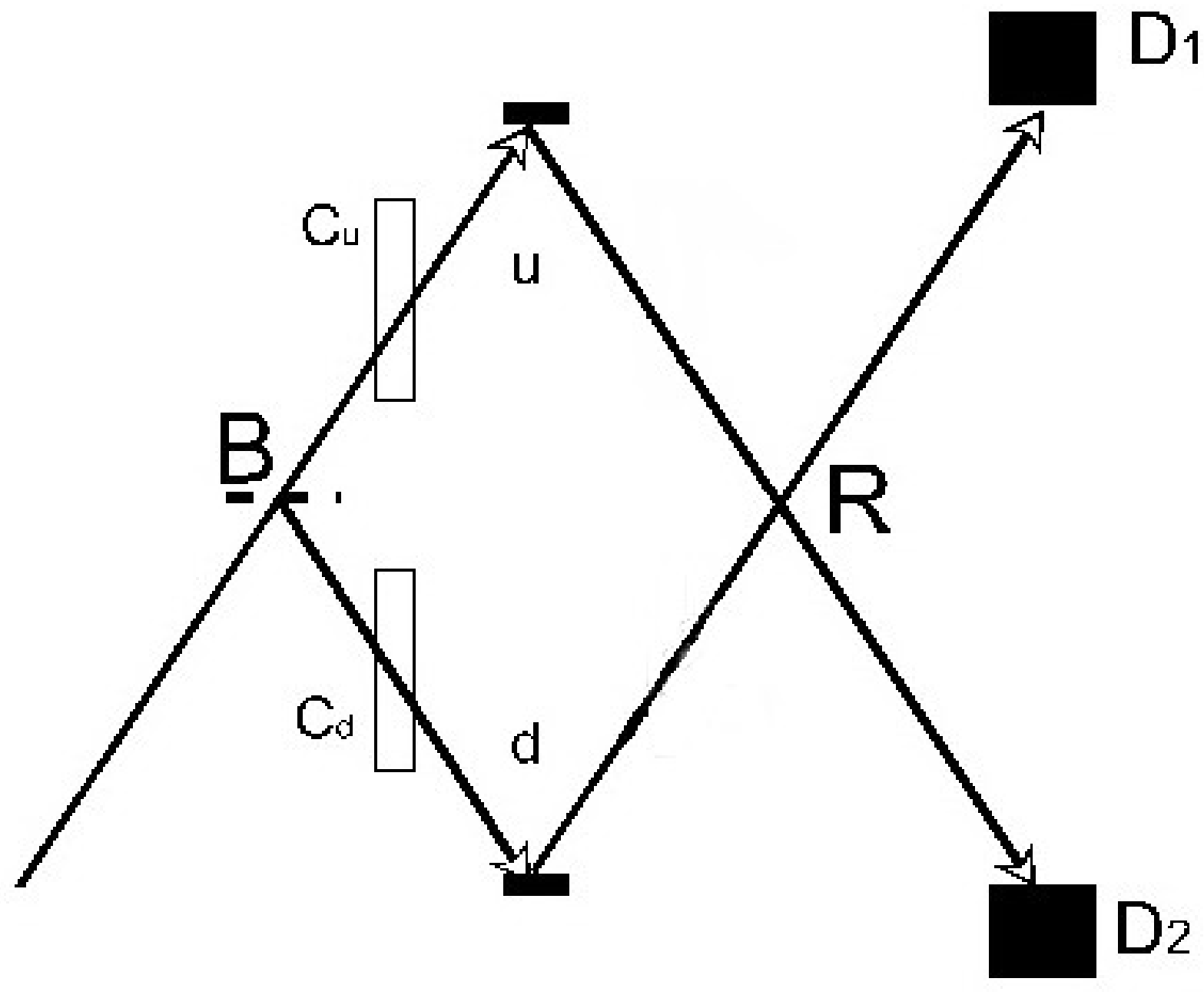}{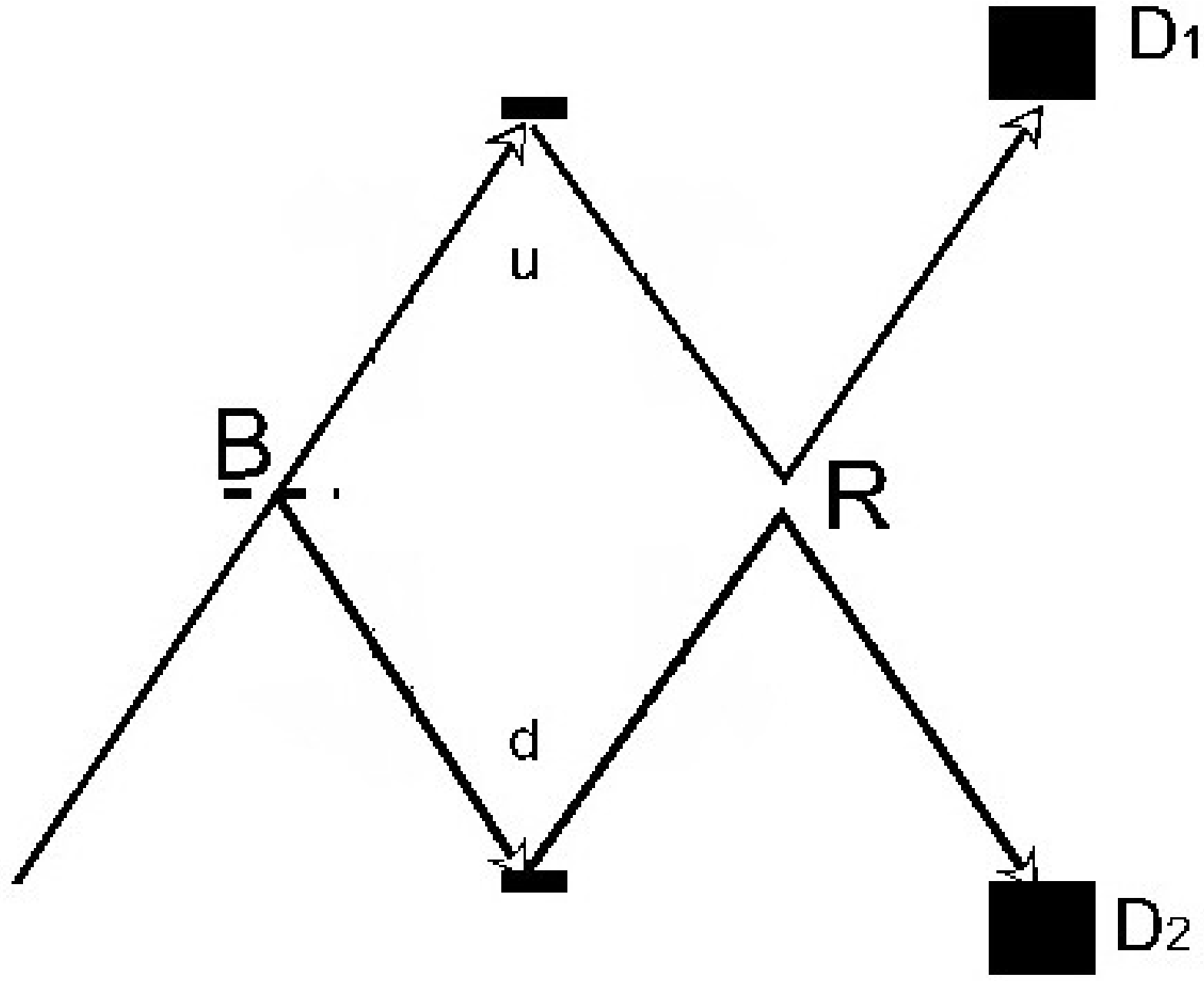} {"Which Path": measured(left); not
measured (right)}{interb}

We will not revisit these arguments here but simply note that,
when a measurement of the path is made while the particle is in
the interferometer (by detectors $C_u$ and $C_d$)
 the Bohm trajectories follow
the paths like Figure \ref{fg:interb}(left).  In the absence of
such a measurement, the Bohm trajectories for the particle are
deflected in the interference region $R$ (Figure
\ref{fg:interb}(right)) so 'fooling' the delayed choice of 'which
path' made by detectors at $D_1$ and $D_2$. \footnote{For a
critique of the 'one bit detectors' critical to the analysis of
\cite{ESSW92} see \cite[Chapter 3]{Mar02}}

For the density matrix we will be considering the same
experimental arrangement, but the atomic state entering the arms
of the interferometer (after the beam splitter region $B$)
 is the mixed state
\begin{equation}
\frac12 \left(\proj{\phi_u(x,t_1)}+\proj{\phi_d(x,t_1)}\right)
\end{equation}
where $\phi_u$ represents a pure state in the upper arm of the
interferometer and $\phi_d$ is a pure state in the lower arm. No
interference effects are expected in the region $R$.

We will describe the Bohm trajectories for this in the cases
where:
\begin{enumerate}
\item The mixed state is a physically real density matrix
$\varrho$;
\item The mixed state is a statistical mixture $\rho$;
\item The mixed state is a physically real density matrix, and a
measurement of the atomic location is performed while the atom is
in the interferometer.
\end{enumerate}

\subsection{Physically real density matrix}
While the atom is in the arms of the interferometer, the
wavepacket corresponding to \proj{\phi_u} and that corresponding
to \proj{\phi_d} are superorthogonal. The trajectories in the arms
of the interferometer are much as we would expect. However, when
the atomic trajectory enters the region $R$, both wavepackets
start to overlap. The previously passive information in the
wavepacket from the other arm of the interferometer becomes active
again.

No interference fringes occur in the region $R$, and if phase
shifters are placed in the arms of the interferometer, their
settings have no effect upon the trajectories\footnote{To observe
interference fringes we would need a density matrix that
diagonalises in a basis that includes non-isotropic superpositions
of \ket{\phi_u} and \ket{\phi_d}.}. However, the trajectories do
change in $R$. The symmetry of the arrangement, and the
'no-crossing principle' for the flow lines in a probability
current, ensures that no actual trajectories can cross the center
of the region $R$. The Bohm trajectories follow the 'surrealistic'
paths similar to those in Figure \ref{fg:interb}(right), even in
the absence of phase coherence between the two arms of the
interferometer.

\subsection{Statistical Ensemble}
We have seen that, even in the absence of phase coherence, the
Bohm trajectories for the density matrix show the surrealistic
behaviour. Does this represent an unacceptable flaw in the model?
To answer this, we now consider the situation where the density
matrix is a statistical ensemble of pure states. This situation
should more properly be described, for the point of view of the
Bohm approach, as a finite assembly\footnote{The term "assembly"
is taken from \cite{Per93}.}
 of many individual systems.

First consider the assembly
\begin{eqnarray}
\rho_1&\equiv& \proj{\phi_{a_1}}\otimes \proj{\phi_{a_2}}\otimes \proj{\phi_{a_3}}\ldots \nonumber \\
&\equiv&\Pi_i \proj{\phi_{a_i}}
\end{eqnarray}
where $a_i=u$ or $d$ with a probability of one-half. As the
assembly consists entirely of product states, the behaviour in
each case is independant of the other cases.

If the state is \proj{\phi_u}, then the trajectories pass down the
u-branch, and go through the interference region without
deflection. Similarly, systems in the \proj{\phi_d} state pass
down the d-branch and are undeflected at $R$. These trajectories
are what we would expect from an incoherent mixture.

However, now let us consider the assembly
\begin{equation}
\rho_2 \equiv \Pi_i \proj{\phi_{b_i}}
\end{equation}
where $b_i=+$ or $-$ occur with equal probability and
\begin{eqnarray}
\ket{\phi_{+}} &=& \frac{1}{\sqrt2}
    \left(\ket{\phi_u}+\ket{\phi_d}\right) \\
\ket{\phi_{-}} &=& \frac{1}{\sqrt2}
    \left(\ket{\phi_u}-\ket{\phi_d}\right).
\end{eqnarray}
This forms exactly the same statistical ensemble. Now, however, in
each individual case there will be interference effects within the
region $R$, it is just that the combination of these effects will
cancel out over the ensemble. If we were to measure the state in
the $(+,-)$ basis, then we would be able to correlate the
measurements of this to the location of the atom on the screen and
exhibit the interference fringes. The Bohm trajectories for the
assembly $\rho_2$ all reflect in the region $R$ and display the
supposed 'surrealistic' behaviour.

There are no observable consequences of the choice of the
different assemblies to construct the statistical
ensemble\footnote{It is interesting to note that if we were to
measure the assembly $\rho_1$ in the $(+,-)$ basis we would still
obtain interference fringes!}. Consequently, if we are only given
the density matrix of a statistical ensemble, we are unable to say
which assembly it is constructed from and cannot simply assume
that the underlying Bohm trajectories will follow the pattern in
Figure \ref{fg:interf}. It is only legitimate to assume the
trajectories will pass through the interference region undeflected
if we know we have an assembly of \ket{\phi_u} and \ket{\phi_d}
states, in which case the Bohm trajectories agree. Thus we
conclude the behaviour of the trajectories for the physically real
density matrix cannot be ruled out as unacceptable on these
grounds.

\subsection{Measuring the path}
Finally, we consider what happens when we have the physically real
density matrix
\begin{equation}
\varrho=\frac12
    \left(\proj{\phi_u(x,t_1)}+\proj{\phi_d(x,t_1)}\right)
\end{equation}
and we include a conventional measuring device in the u-path. The
measuring device starts in the state \ket{\xi_0}. If the atom is
in the state \ket{\phi_u}, the measuring device moves into the
state \ket{\xi_1}. The states \ket{\xi_0} and \ket{\xi_1} are
superorthogonal.

If we now apply the interaction to the initial state
\begin{equation}
\varrho \otimes \proj{\xi_0}
\end{equation}
the system becomes the correlated density matrix
\begin{equation}
\frac12
    \left(\proj{\phi_u\xi_1}+\proj{\phi_d\xi_0}\right).
\end{equation}
As we saw above, as the measuring device states are
superorthogonal, the system behaves exactly as if it were the
statistical ensemble. This is true even when the atomic states
enter the region $R$. The Bohm trajectories of the atom pass
undeflected through in the manner of Figure \ref{fg:interb}(left).

We conclude that the Bohm trajectories for the density matrix
cannot be considered any more or less acceptable than the
trajectories for the pure states.

\section{CONCLUSION}\label{conclusion}
By extending the de Broglie-Bohm interpretation to cover density
matrices, we showed it was possible to consistently treat the
density matrix as a property, not of an ensemble, but of an
individual system.

It is worth asking why Bell seemed to suggest that this was not
possible.  An examination of \cite{Bel80} shows that the density
matrix Bell considered arose, not as a fundamental description of
an individual system, but through tracing over part of an
entangled system in an overall pure state, as in Equation
\ref{eq:reduced}.

Bell considers a pure state that interacts with a measuring
device, in an arrangement similar to Figure \ref{fg:interb}(left).
After the detectors have interacted with the atomic beam, the full
state is:
\begin{equation}
\frac{1}{\sqrt{2}}\left(\ket{\phi_u\xi_1}+\ket{\phi_d\xi_0}\right).
\end{equation}
Tracing over the detector states $\xi_i$ leaves the density matrix
\begin{equation}
\rho=\frac12
\left(\proj{\phi_u}+\proj{\phi_d}\right)\label{eq:bellrho}.
\end{equation}
Bell wishes to construct trajectory solutions which resemble
Figure \ref{fg:interb}(left) based on the density matrix for the
atomic beam alone.  This fails, as the entangled state
trajectories are non-locally dependant upon the location of beable
of the measuring device.

The symmetry of Equation \ref{eq:bellrho} and the 'no-crossing'
principle ensure that a trajectory solution based upon the reduced
density matrix alone must produce tracks as in Figure
\ref{fg:interb}(right).  It is only the superorthogonality of the
detector states $\ket{\xi_i}$ that allows the atomic beam states
to act as if they are a statistical ensemble of $\ket{\phi_i}$
states and cross over in the interference region\footnote{It is
almost certainly the case that Bell's motivation was quite
different to this paper.  A possible concern would be the
complication of constructing trajectory solutions for many-body
entangled states.  If a reduced density matrix could be used, it
would not be necessary to include a description of all the degrees
of freedom for all subsystems.  Unfortunately this is not
possible.}.

However, the density matrix we have considered in this paper does
not arise either as a reduced density matrix, or as a statistical
ensemble of pure states.  It is the complete description of an
individual system.  For this, we have shown that a consistent and
coherent trajectory interpretation in the manner of the de
Broglie-Bohm interpretation is possible.

\bibliographystyle{alpha}
\bibliography{paper6}

\end{document}